\documentclass[useAMS,usenatbib]{mn2e}
\usepackage{graphicx}
\title[SMBH Mass Regulated by Host Galaxy Morphology]
{Supermassive Black Hole Mass Regulated by Host Galaxy Morphology}

\author[Y. Watabe et al.]
{Y. Watabe $^{1}$\thanks{E-mail: watabe@arcetri.astro.it},  
N. Kawakatu $^{2}$, 
M. Imanishi $^{2}$, 
T. T. Takeuchi $^{3}$
\\
$^{1}$INAF-Osservatorio Astrofisico di Arcetri, L.go E. Fermi 5, 50125 Firenze, Italy\\
$^{2}$National Astronomical Observatory of Japan, 2-21-1 Osawa, Mitaka, Tokyo 181-8588, Japan\\
$^{3}$Institute for Advanced Research, Nagoya University, Furo-cho, Chikusa-ku, Nagoya 464-8601, Japan\\
}

\begin{document}

\date{}


\maketitle

\label{firstpage}

\begin{abstract}
We investigated the relationship between supermassive black hole (SMBH) mass and host starburst luminosity in Seyfert galaxies and Palomar-Green QSOs, focusing on the host galaxy morphology. Host starburst luminosity was derived from the 11.3 $\micron$ polycyclic aromatic hydrocarbon luminosity. We found that the SMBH masses of elliptical-dominated host galaxies are more massive than those of disk-dominated host galaxies statistically. 
We also found that the SMBH masses of disk-dominated host galaxies seem to be suppressed even under increasing starburst luminosity. These findings imply that final SMBH mass is strongly regulated by host galaxy morphology. This can be understood by considering the radiation drag model as the SMBH growth mechanism, taking into account the radiation efficiency of the host galaxy.
\end{abstract}


\section{Introduction}
Recent observations have found that the mass of supermassive black holes (SMBH) in galactic centers correlates strongly with bulge mass in both active and inactive galaxies \citep{KG01,MF01,Tr02,MH03,Ba05}. Also, it has been revealed that SMBH mass does not correlate with galaxy disk mass; in fact, the SMBH--disk mass ratio is much smaller than the SMBH--bulge mass ratio \citep{Sa00,KG01}. These observational findings suggest that SMBH formation is strongly connected to the bulge component, not the disk component. However, the physical mechanism of SMBH formation that can lead to these observational results has remained unclear. 

Investigations of starbursts around active galactic nuclei (AGNs) may prove pertinent to this question. Observations of polycyclic aromatic hydrocarbon (PAH) emission have gradually revealed evidence of starburst phenomena around AGNs and a connection between AGN activity and starbursts \citep{Im02,Im03,RV03,IW04,Sc06,Ma07,Ne07,Sh07,Wa08,Lu08}. This means that the mass accretion process onto the SMBH, i.e., the SMBH growth mechanisms, could be closely connected with starburst phenomena.

\begin{table*}
\begin{tabular}{lrcc|lrcc}
\hline
name & 
$\log L_{11.3 \micron}$ & 
$\log M_{\rm BH}$ &
references&
name & 
$\log L_{11.3 \micron}$ & 
$\log M_{\rm BH}$ &
references\\
& 
$[L_{\odot}]$ & 
$[M_{\odot}]$ &
& 
& 
$[L_{\odot}]$ &
$[M_{\odot}]$ &\\
\hline
--- Elliptical --- &&&&--- Bulge + Disk ---&&&\\
PG 0923$+$201  & $<$ 8.77 & 8.91 &1, 3, 4, 8&
PG 0007$+$106 & 8.30 & 8.25 &2, 8\\
PG 1004$+$130  & 8.84 & 9.07 &1, 3&
PG 0026$+$129 & $<$ 8.11 & 7.81 &1, 6\\
PG 1121$+$422  & $<$ 8.43 & 7.91$^\dagger$ &6&
PG 0043$+$039 & $<$ 8.92 & 9.19 &4, 7\\
PG 1202$+$281  & 8.74 & 8.26 &4, 8&
PG 0050$+$124 & 8.69 & 7.22 &1, 2\\
PG 1216$+$069 & $<$ 8.56 & 9.13 &4, 7&
PG 0052$+$251 & 8.61 & 8.68 &3, 4, 8, 9\\
PG 1302$-$102  & $<$ 8.83 & 8.27 &1, 4, 8&
PG 0157$+$001 & 9.55 & 8.06$^\dagger$ &2, 3, 9\\
PG 1307$+$085 & $<$ 8.75 & 7.82 &1, 4, 8&
PG 0804$+$761 & $<$ 7.98 & 8.27 &1\\
PG 1322$+$659 & 8.49 & 8.12 &6&
PG 0838$+$770 & $<$ 8.31 & 7.95 &1\\
PG 1427$+$480 & $<$ 7.94 & 7.96 &6&
PG 0953$+$414 & $<$ 8.82 & 8.54 &3, 4, \\
PG 1435$-$067  & $<$ 8.20 & 8.20 &1&
PG 1012$+$008 & $<$ 8.25 & 7.76 &3, 4, 5, 9\\
PG 1613$+$658 & 8.83 & 8.95 &6&
PG 1048$+$342 & $<$ 7.79 & 8.21$^\dagger$ &6\\
PG 1617$+$175 & $<$ 8.49 & 8.66$^\dagger$ &1&
PG 1116$+$215 & $<$ 8.64 & 8.18 &1, 4, 8\\
PG 1704$+$608 & $<$ 9.02 & 7.84 &5&
PG 1119$+$120 & 7.97 & 7.16 &1,2 \\
PG 2214$+$139 & $<$ 7.76 & 8.38$^\dagger$ &1&
PG 1126$-$041 & $<$ 8.37 & 7.53$^\dagger$ &1, 2\\
PG 2349$-$014 & 8.84 & 8.75 &9&
PG 1151$+$117 &  $<$ 8.92 & 8.40$^\dagger$ &6\\
--- Disk ---&&&&
PG 1229$+$204 & 7.86 & 7.89 &1, 2, 4\\
PG 0844$+$349 & 7.87 & 7.65 &1&
PG 1309$+$355 & $<$ 8.84 & 8.17 &1, 4, 8\\
PG 1001$+$054 & 8.38 & 7.59$^\dagger$ &1, 5&
PG 1354$+$213 & $<$ 8.54 & 8.50 &6, 7\\
PG 1211$+$143 & $<$ 7.68 & 7.84 &9&
PG 1444$+$407 & $<$ 8.82 & 8.19 &4, 8\\
PG 1352$+$183 & $<$ 9.51 & 8.23 &6&
PG 2130$+$099 & 8.01 & 7.64 &1, 2\\
PG 1402$+$261 & $<$ 8.51 & 7.26 &4, 5, 8&
PG 2233$+$134 & $<$ 9.02 & 8.00$^\dagger$ &6, 7\\
PG 1440$+$356 & 8.84 & 7.26 &1, 9&&&&\\
PG 1543$+$489 & $<$ 9.46 & 7.99$^\dagger$ &7&&&&\\
\hline
\end{tabular}
\caption{QSOs properties. Column 1, 5: Object name. Col. 2, 6: 11.3 $\micron$ PAH luminosity. Col. 3, 7: SMBH mass estimated by Kawakatu et al. (2007). $\dagger$: the data are from Vestergaard et al. (2006). Col. 4, 8: References of host galaxy morphology; [1] \citet{Gu06}; [2] \citet{Ve06}; [3] \citet{Du03}; [4] \citet{Ha02}; [5] \citet{Ma01}; [6] \citet{MM01}; [7] \citet{Pe01}; [8] \citet{Ba97}; [9] \citet{Ta96}.
\label{tb1}
}
\end{table*}

\citet{Um01} and \citet{KU02} suggested that the physical mechanism of the link between the SMBH and bulge formation may be the radiation drag effect (the Poynting-Robertson effect) from bulge starbursts. Especially, \citet{KU04} showed that the radiation efficiency differs for starbursts in the bulge compared to the disk. The bulge is round, providing high radiation efficiency. Disk starbursts are less efficient than bulge starbursts in the same starburst luminosity range due to photon escape from the disk surface and edge-on opacity consideration, so the final SMBH mass of a host galaxy with a disk starburst cannot be large. Thus, to understand SMBH growth mechanisms, it may be necessary to consider where the starbursts occur in their host galaxies. In order to understand the radiation effects from the host starburst and confirm the radiation drag model, we must investigate the relationship between SMBH mass and host starburst activity, focusing on host galaxy morphology.

To date, the morphology of Seyfert galaxies (low luminosity AGNs) have been well studied and is almost all spiral \citep[e.g.,][]{MR95,HM99}. 
Moreover, the starbursts in the galaxy disk have been investigated \citep{Wa08}. For QSOs (high luminosity AGNs), although it has been difficult to determine their morphology due to their luminous nuclei, recent high-resolution imaging observations in the optical and near infrared, and those exploiting adaptive optics (AO) have gradually revealed their morphology in detail \citep{Ta96,Ba97,Pe01,MM01,Ma01,Ha02,Du03,Ve06,Gu06}. They found not only the elliptical  (which is equivalent to the bulge, classified as the spheroid) component but also the prominent disk component in the host of lower luminosity QSOs and radio-quiet QSOs \citep{Du03}. 
Also, we now know that QSO host galaxies are gas-rich (which is not normal for elliptical galaxies) \citep{Ev01,Ev06,Sc03} and that  starbursts occur in the host galaxies \citep{Ha03,Ba06,Sc06,Ma07,Ne07,Sh07,Lu08}.

In this letter, to clarify the SMBH growth mechanism that satisfies the observational results, we investigated the SMBH mass--host starburst connection, taking into account the host galaxy morphology for Seyfert galaxies and QSOs. Throughout the paper, we adopted $H_{0} = 80\, \rm km\,s^{-1} Mpc^{-1}$, $\Omega_{M}=0.3$, and $\Omega_{\Lambda}=0.7$.

\section[]{Data and Analysis}

\subsection[]{Sample data}

To investigate the effect of starbursts and host galaxy morphology on the final SMBH mass, we selected AGNs based on estimated SMBH mass, PAH emission, and morphology. The samples in this paper are Palomar-Green (PG) QSOs, selected at $B$ band to have blue $U-B$ color 
 (Schmidt \& Green 1983). Since the B- and U-band emission of PG-QSOs are dominated 
by the AGN, there are no biases for the presence of starbursts.

Seyfert galaxies in the CfA \citep{HB92} and 12 $\mu$m \citep{Ru93} samples, selected on the basis of their host galaxy magnitudes and $IRAS$ 12 $\mu$m fluxes, respectively. 
These samples do not also include some biases for the presence of starbursts.

\subsection[]{Black hole mass}

To estimate SMBH mass, $M_{\rm BH}$, of PG-QSOs and Seyfert 1 galaxies, we used a method 
based on the reasonable assumption that the motion of ionized gas clouds around the SMBH is dominated by the gravitational force, and that the clouds within the broad-line region (BLR) are virialized \citep[e.g.,][]{PW99}. The velocity dispersion $v$ can be estimated from the FWHM of H$\beta$ broad line emission $v=fv_{\rm FWHM}$ 
by assuming the isotropic spherical virial coefficient, $f=\sqrt{3}/2$ (Netzer 1990). We selected the H$\beta$ because H$\beta$ lines radiate by simple photoionization mechanisms and that their line profiles reflect gravitational potential of the H$\beta$ emission region.
Adopting an empirical relationship \citep{Ka00} between the size of the BLR and the rest-frame optical continuum luminosity, 
$\lambda L_{\lambda} (5100 \AA)$, and using reverberation mapping, we obtain the following formula:
\begin{equation}
M_{\rm BH} = 4.9 \times 10^{6} 
\left[ \frac{\lambda L_{\lambda} (5100 \rm \AA)}{10^{44}\, \rm{ergs\, s^{-1}}} \right]^{0.70} \left( \frac{v_{\rm FWHM}}{10^{3}\, \rm{km\, s^{-1}}} \right)^{2} M_{\odot}.
\end{equation}
For the error of SMBH mass, McGill et al. 2008 have compared 12 formulae taken from the literature, showing that SMBH mass estimates can differ on average $0.13 \pm 0.05$ or $0.38 \pm 0.05$ dex in the case of the same or different virial coefficient, respectively. 
For Seyfert 2 galaxies, we used SMBH mass data \citep{BG07} whose mean error was within a factor of 1.6. These were estimated from the SMBH mass--stellar velocity dispersion relation \citep{Tr02}.

\begin{figure*}
\includegraphics[width=8cm]{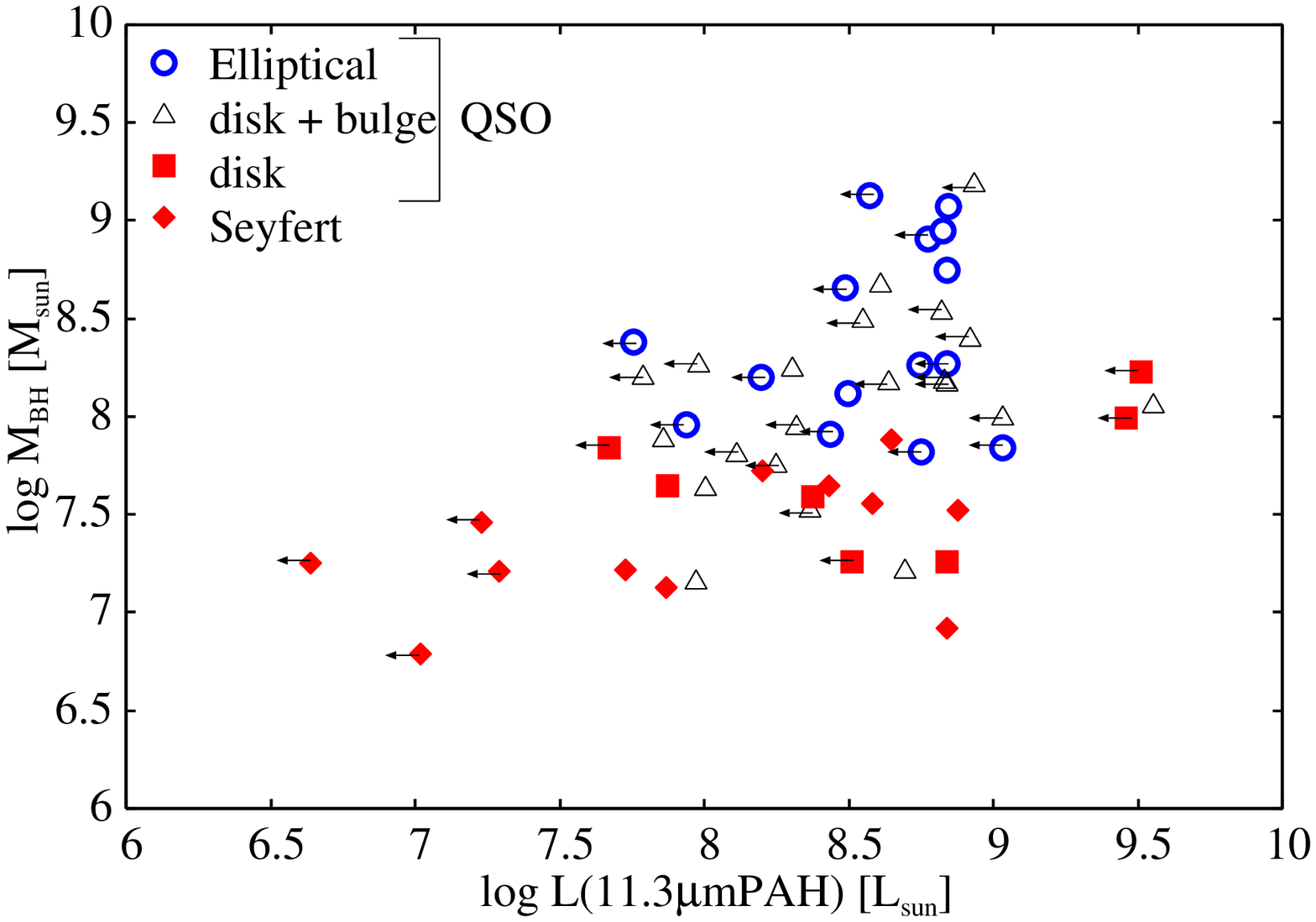}
\includegraphics[width=8cm]{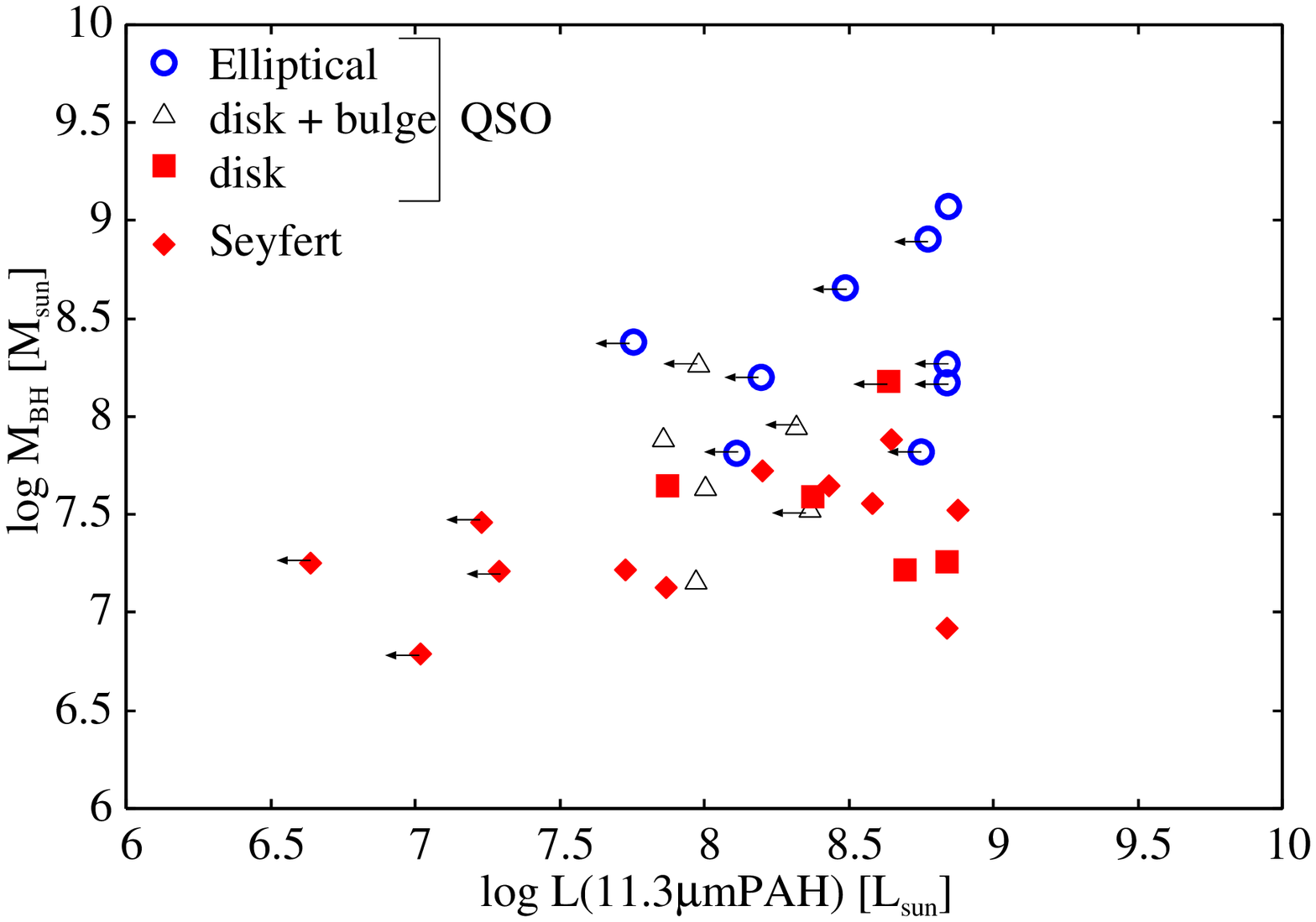}
\caption{The 11.3 $\mu$m PAH luminosity vs. SMBH mass for a given the host morphology. The morphology classification of PG-QSOs host galaxy is used Table 1 ({\it left}) and only \citet{Gu06} ({\it right}). {\it Open circle, open triangle,} and {\it filled square}: elliptical-dominated, bulge+disk, and disk-dominated PG-QSOs, respectively. {\it Filled diamond}: Seyfert galaxies. } 
\label{fig1}
\end{figure*}

\subsection[]{Host galaxy starburst luminosity}

Since the PAH molecules are excited by far-UV photons in the photo-dissociation region around the HII region, and strong PAH emission is often observed from even a weak starburst \citep{Im02}, we can use PAH emission as an indicator of starburst activity. We used the PAH emission estimated by \citet{Wa08} for Seyfert galaxies (6.2, 7.7, and 11.3 $\mu$m) and \citet{Sh07} for PG-QSOs (7.7, and 11.3 $\mu$m). We select the 11.3 $\mu$m PAH emission. The 7.7 $\mu$m PAH emission is sometimes affected by the broad and strong 9.7 $\mu$m silicate absorption, especially in Seyfert 2 galaxies. Thus, it could be difficult to distinguish between 7.7 $\mu$m PAH emission and 9.7 $\mu$m silicate absorption. PAH emission was obtained with the {\it Spitzer Space Telescope} Infrared Spectrograph \citep[IRS;][]{Ho04}  \citep{We04}. For Seyfert galaxies, since the PAH emission was observed with the slit-scan mode (PID 3269, PI: J. Gallimore), the entire host galaxy regions of the Seyfert galaxies were covered. For PG-QSOs, PAH emission was obtained by the slit width of Short-Low (SL) (SL1: 3$\arcsec$.7, SL2: 3$\arcsec$.6) and Long-Low(LL) (LL1: 10$\arcsec$.7, LL2: 10$\arcsec$.5) modules \citep[for PID and PI, see][]{Sh07}. The SL slit width is roughly comparable to the effective radius of PG-QSO host galaxies \citep{Gu06} and several hundred-parsec to kiloparsec-scale starbursts have been considered as the origin of the far-infrared radiation in PG-QSOs \citep{Ha03,Ba06,Ne07}. Thus, we consider that these slit observations cover almost all starburst activity in PG-QSOs host galaxies.

\subsection{Host galaxy morphology}

We used host galaxy morphology classifications for PG-QSOs based on the literature, evaluated by 2D $\chi^2$ fitting of the obtained image (references listed in Table 1). We carefully checked the morphology classifications and defined a host galaxy as elliptical- or disk-dominated only in cases for which the literature listed said galaxy as only elliptical or disk, respectively. We also defined a bulge + disk host galaxy in cases where a 2D fit favored a two-component (bulge + disk) model, or in cases where the morphology decision given in the literature varied (see Table 1). We also used a homogeneous morphology classification criterion to check our results, by separately examining objects classified only by \citet{Gu06}. This sample consisted of a number of PG-QSOs (20 objects) investigated by near-infrared AO imaging with the Gemini-N and Subaru Telescope. For the Seyfert galaxies, since their hosts are almost all spiral galaxies \citep{MR95,HM99}, we assumed that their morphologies were disk-dominated.

\section{Results: SMBH mass and host starburst geometry relationship}

We plotted the 11.3 $\mu$m PAH luminosity and SMBH mass specifying the host galaxy morphology in  Figure \ref{fig1}. The left and the right panels of this figure show that the morphology classification is used Table 1 and only \citet{Gu06}, respectively.

We applied detailed statistical tests about the difference of these distributions (Takeuchi et al. 2009 in prep.) . To summarize, the SMBH mass distributions of the elliptical- and disk-dominated host galaxies are significantly different. However, the difference of the PAH luminosity is not very clear. (We showed this statistical analysis in Appendix.) 
Averaged SMBH mass and its dispersions are $<\log M_{\rm BH}/M_{\odot}> = 8.42\, (\pm 0.44) $ and $7.48\, (\pm 0.36)$ for the elliptical- and disk-dominated host galaxies, respectively. 
Also, bulge + disk host galaxies are distributed in both the elliptical- and the disk-dominated host galaxy regions. In particular, for the disk-dominated host galaxy, although the PAH luminosity increases by about three orders of magnitude, SMBH mass increases by only about one order. These results indicate that the final SMBH mass is strongly connected with host galaxy morphology; the SMBH of a disk-dominated host galaxy is suppressed, while a more massive SMBH can form in an elliptical-dominated host galaxy.
In order to remove the differences in observations and morphology classification methods, we checked our findings using data from \citet{Gu06} only in the right panel of Figure 1.  Although the sample size decreases, the tendency of our results does not change; $<\log M_{\rm BH}/M_{\odot}> = 8.40 \,(\pm 0.41) $ and $7.42\, (\pm 0.34)$ for the elliptical- and disk-dominated host galaxies, respectively.

\section{Discussion: AGN Fueling Mechanism}
To interpret our findings, we must consider the SMBH growth mechanism including both the host starburst effects and the host galaxy morphology. Although galaxy mergers \citep[e.g.,][]{He89} and stellar bars \citep[e.g.,][]{No88} have also been considered as SMBH growth mechanism candidates, the relationship between final SMBH mass and these mechanisms is still unknown. Therefore, it is difficult to explain the difference in SMBH mass for the same starburst luminosity range using these mechanisms. Thus, in order to relate the host starburst and host galaxy morphology with SMBH formation, we focused on the radiation-hydrodynamic effect from the host starburst. The radiation drag is a relativistic effect known as the Poynting-Robertson effect. It is a possible mechanism for extracting angular momentum from the gas and driving SMBH mass accretion \citep{Um97,Um01,KU02}. Final SMBH mass is connected with the absorption efficiency of the amount of radiation energy from the starburst. This radiation drag efficiency is strongly affected by host geometry \citep{Um97,Um98,Oh99}. \citet{KU04} explored the  possibility that SMBH mass of a disk-dominated host galaxy could be one to two orders of magnitude smaller than that of an elliptical-dominated host galaxy due to effects of geometrical dilution and opacity. If an elliptical-dominated host galaxy begins with starbursts in a highly inhomogeneous and optically thick interstellar medium \citep[e.g.,][]{Sa88,Go97}, radiation drag could effectively work to extract the angular momentum \citep{KU02}. In contrast, a large number of photons escape from the disk surface of a disk-dominated host galaxy. Also, radiation from a disk-starburst is shielded due to edge-on optical depth. Therefore, from a radiation-hydrodynamic point of view, final SMBH mass could be strongly connected to starburst location. 
Also, there seems to be no possibility that the radiation efficiency is equivalent for disk- and elliptical-dominated host galaxies and the SMBH mass of a disk-dominated host galaxy is currently small, but will grow up in the future due to SMBH growth delay. It's because SMBH masses of the inactive spiral galaxies are comparable to those of the disk-dominated host galaxies with AGNs and follow the same SMBH--bulge mass relation \citep{Pa07}. Therefore, in the case that all inactive spiral galaxies have passed the AGN phase \citep{Ma04}, the final SMBH masses of the disk-dominated host galaxies with AGNs could not reach those of the elliptical-dominated host galaxies. Thus, our result flows naturally from a  radiation drag model that includes the radiation efficiency due to host galaxy morphology. 

\section{Summary}
We investigated the SMBH mass--host starburst connection, taking into account the host galaxy morphology for Seyfert galaxies and PG-QSOs. We checked the statistical test about the difference of these distributions. As a result, we found that host galaxy morphology may strongly regulate final SMBH mass, as the SMBH masses of elliptical-dominated host galaxies were more massive than those of disk-dominated host galaxies. Also, the SMBHs of disk-dominated host galaxies showed suppressed mass even in the case of increased starburst luminosity. These findings indicate that the SMBH growth mechanism is strongly connected to radiation efficiency dependent on the geometry of the host starbursts. 

\section*{Acknowledgments}
We thank the anonymous referee for valuable comments.

\section*{Appendix}
Here we explained the statistical tests. (Further details will be found in Takeuchi et al. (2009 in prep.).)
First, we tested if the vertical marginal distributions of Figure~1
are the same. This is a well-known statistical problem which is properly 
addressed by the Kolmogorov-Smirnov two sample test
\citep[e.g.,][]{Si56}.
The test statistic is the distance between the two distribution functions (DFs):
$  D \equiv \max \left| F_{a}(M_{\rm BH}) - F_{b}(M_{\rm BH}) \right| \; $,
where the subscripts $a$ and $b$ represent that they are the DFs of 
the sample in the left and right panel of Figure 1, respectively.
Then, it is known that a quantity
\begin{eqnarray}\label{eq:KM_chi2}
  \chi^2 \equiv \frac{4 D^2 n_{\rm ell} n_{\rm disk}}{n_{\rm ell}+n_{\rm disk}}
\end{eqnarray}
obeys the $\chi^2$-distribution with a degree of freedom (dof) of 2, 
under the null hypothesis that these are drawn from the same DF.
The distances $D$ from the sample of 
$a$ and $b$ are
0.818 and 0.882.
Then, the probabilities having the $\chi^2$ values [eq.~(\ref{eq:KM_chi2})]
corresponding to these values are $5 \times 10^{-6}$ and $0.0001$, 
respectively.
Thus, in both cases, the null hypothesis is very clearly rejected.

Next, we examine the horizontal marginal distribution of the samples.
Along with the abscissa in Figure~1, we have a significant number of
upper limits both in elliptical and disk galaxy samples.
In this case, usual two sample tests are no longer valid.
We first estimate DFs of the PAH luminosity, $F(L_{\rm PAH})$ of both samples 
by the Kaplan-Meier (KM) estimator \citep{KM58,KP02}.
The KM estimator is designed to obtain the so-called survival 
function (or survivor function), $S(t)$, of a given sample including 
right censored data (i.e., data with lower limits).
A survival function is related to the DF by the following relation: 
 $S(t) = 1 - F(t)$.

{}To construct this estimator, we first mathematically formulate
the current problem according to \citet{FN85}.
Since most of the astronomical observational data are
selected with a certain detection limit, usually a sample includes
upper limits, instead of lower limits. 
In statistical terminology, lower-limit data are called ``left-censored''.
Let $T_1^L, T_2^L, \dots, T_n^L$ denote measurements, where the
superscript $L$ means ``left''.
If $T_i^L < A_i$, where $A_i$ is its upper limit, then we cannot
obtain the exact value of $T_i$ but the upper limit $A_i$.
The information available is a combination of 
$\{X_i^L, \delta_i^L\} = \{T_i, 1\}$ and $\{A_i, 0\}$ for $T_i^L \ge A_i$ and $T_i^L < A_i$, respectively.
Suppose that $T_i^L$ is a random sample drawn from a distribution
$F^L(t) = P(T_i \le t)$ where $P(E)$ denotes the probability of 
an event $E$.
If $\{A_i\}_{i = 1, 2, \dots, n}$ are mutually independent, identically
distributed (referred to as IID) and independent of the true measurements
$\{T_i^L\}_{i = 1, 2, \dots, n}$, this statistical model is called
random censorship.
We can easily transform left-censored data to right-censored 
ones by setting a constant $M$; $T_i = M - T_i^L$, $X_i = M - X_i^L$, $C_i = M - A_i^L$, $\delta_i = \delta_i^L$.
Then, $\{X_i,\delta_i\}_{i = 1, 2, \dots, n}$ represent right-censored
data.
The KM estimator has the following form:
\begin{eqnarray}
  \hat{S}(t) = 
    \left\{
      \begin{array}{ll}
        \prod_{j|x'_{(j)} < t} 
        \left( 1 - \frac{d_j}{n_j}\right)^{\delta_{(j)}} & t > x'_{(j)}\\
        \\
        1 & t \leq x'_{(j)}
      \end{array}
    \right. \;,
\end{eqnarray}
where $x'_{(j)}$ denote distinct, ordered observed values in which ties
are identified, and $n_j = \#\left\{ k; x_k \ge x'_{(j)} \right\}$, $d_j = \#\left\{ k; x_k = x'_{(j)} \right\}$. 
A detailed but comprehensive derivation is found in, e.g., 
\citet{FN85} or \citet{KP02}.

Once we get the KM estimate of the survival function, $\hat{S}(t)$, we
can convert it into the estimate of the DF, $\hat{F}(t)$, as
\begin{eqnarray}
  F^L(t) &=& P\left(T^L \le t\right)\nonumber \\ 
      &=& P\left(M-T \le t \right) \nonumber \\
   &=& P\left(T \le M-t \right) \nonumber \\
    &=& S(M-t) \;.
\end{eqnarray}
Then, 
\begin{eqnarray}
  \hat{F}^L(t) = \hat{S}(M-t) \;.
\end{eqnarray}
In this analysis, we set $M=0$, i.e., we made a flip of a sign 
$t \rightarrow -t$.

In order to see if these two DFs are different, we should perform 
a statistical test.
Though there are a few methods to do so, we adopt the Mantel-Haenszel
(MH) two-sample test (also often referred to as the logrank test, 
but the latter denotes some variations).
The derivation is shown in, e.g., \citet{KP02}.
The MH test makes use of the values as follows:
first we sort all the galaxy sample along with the (minus logarithmic)
luminosity, $- \log {L_{\rm PAH}}_{,j}$. 
Then, we define mean numbers of galaxies at each luminosity 
$- \log {L_{\rm PAH}}_{,j}$ for both samples,
  $m_{k,j} \equiv d_{j} \frac{n_{k,j}}{n_j}$
where $k$ denotes the label of the two samples, ``ell'' and ``disk'', and
  $n_j \equiv n_{{\rm ell},j} + n_{{\rm disk},j}$.
We also define variance
\begin{eqnarray}
  v_{kk,j} &=& v_{\ell\ell,j} \nonumber \\
    &\equiv& m_{k,j} \left( 1 - \frac{n_{k,j}}{n_j}\right)
    \left( \frac{n_j - d_j}{n_j-1}\right) \nonumber \\
    &=& \frac{n_{k,j}n_{\ell,j}d_j \left(n_j-d_j\right)}{
      \left(n_j\right)^2 \left( n_j -1\right)}
\end{eqnarray}
($k$ and $\ell$ again denote ``ell'' and ``disk''), and 
covariance\footnote{However, in this case, we have only two samples, 
they are equivalent but the signs are different.
We use the variance-covariance matrix when we perform $n$-sample test 
($n > 2$).}; $v_{k\ell,j} = -v_{kk,j}$. 
Consider a deviation vector $\left(\sum_j \Delta_{{\rm ell},j}, 
\sum_j \Delta_{{\rm disk},j} \right) = 
\left( \sum_j d_{{\rm ell},j} - m_{{\rm ell},j} , 
\sum_j d_{{\rm disk},j} - m_{{\rm disk},j} , \right)$.
Since the dof is $2 - 1 = 1$ in the current problem,
$\sum_j \Delta_{{\rm ell},j} = -\sum_j \Delta_{{\rm disk},j}$.
Then, simply we can use a statistic ${\Delta_{k}}^2/v_{kk}$.
If the two samples are drawn from the same DF (this is the null 
hypothesis to be tested), this statistic should obey the 
$\chi^2$-distribution with $\mbox{dof}=1$.
{}From the sample of 
$a$ and $b$, we have $\chi^2= 0.694$
and 0.951, respectively.
Probabilities of having these values under the $\chi^2$-distribution
are 0.405 and 0.329.
Then, if we set the confidence limit of 0.05, we cannot reject the 
null hypothesis, i.e., the two DFs are not significantly different.

\end{document}